
\nopagenumbers
\overfullrule0pt
\def\dots{.~.~.}
\def\rta{\rightarrow}
\def\ifmath#1{\relax\ifmmode #1\else $#1$\fi}
\def\third{\ifmath{{\textstyle{1 \over 3}}}}
\def\ls#1{\ifmath{_{\lower1.5pt\hbox{$\scriptstyle #1$}}}}
\def\lam{\lambda}
\def\mf{m_f}
\def\hsm{\phi^0}
\def\mhsm{m_{\hsm}}
\def\mw{m_W}

\def\hpm{H^{\pm}}
\def\mhpm{m_{\hpm}}
\def\mhl{m_{h^0}}
\def\mhh{m_{H^0}}
\def\mha{m_{A^0}}

\def\ha{A^0}
\def\crr{\crcr\noalign{\vskip .15in}}
\def\SCIPP{\centerline {\it Santa Cruz Institute for Particle Physics}
  \centerline{\it University of California, Santa Cruz, CA 95064}}
\def\DAVIS{\centerline {\it Department of Physics}
  \centerline{\it University of California, Davis, CA 95616}}
\def\MICHIGAN{\centerline {\it Department of Physics}
  \centerline{\it University of Michigan, Ann Arbor, MI 48109}}
\def\BROOKHAVEN{\centerline {\it Physics Department}
  \centerline{\it Brookhaven National Laboratory}
  \centerline{\it Upton, NY 11973}}
\def\doeack{\foot{Work supported, in part, by the Department of Energy.}}

\footline{\hfil\folio\hfil}
\headline{\hfil\null\hfil}
\footline{\hfil\null\hfil}
\phantom{This is a blank page.}
\endpage

\Pubnum={SCIPP-92/58}
\date={December 1992}
\pubtype{}
\titlepage
\baselineskip 0pt
\title{Errata for {\it The Higgs Hunter's Guide}\doeack}
\author{John F. Gunion}
\vskip .1in
\DAVIS
\author{Howard E. Haber}
\vskip .1in
\SCIPP
\author{Gordon L. Kane}
\vskip .1in
\MICHIGAN
\author{Sally Dawson}
\vskip .1in
\BROOKHAVEN

\abstract
\baselineskip 0pt

Errata are given for {\it The Higgs Hunter's Guide}.  These errata should
be applied to the {\it second} printing of the book, dated 1991.
The second printing has already corrected numerous errors and misprints
contained in the originally published 1990 edition.
\endpage
\titlestyle{Errata for {\it The Higgs Hunter's Guide}, Second Printing
(1991).}

\pointbegin
On p.~30, eq.~(2.25) should read:
$$\lim_{\mf\rta\infty}\ I_1(\tau_f,\lam_f)-I_2(\tau_f,\lam_f)=
-\third\,.$$
\point
On p.~62, in eq.~(2.131), the denominator on the right hand side should
read $\sigma_0(\nu_\mu N\rightarrow\mu^-X)$.
\point
On p.~96, eq.~(3.9) is not correct.  To obtain the correct answer,
we must compute the contribution of fig.~(3.7) to $\Delta a_\mu$.
Assuming that the interaction is specified in eq.~(3.7),
where $h$ is now the negatively charged Higgs boson ($H^\pm$),
and $F$ is now a neutral-charged fermion, we find:
$$
\left.\Delta a_{\mu} \right|_{\rm fig.~3.7}
 ={-m_{\mu}^{2} \over 8 \pi^{2}}
\int_{0}^{1} x(1-x)\,dx {\big\{ C_{S}^{2} (x +{m\ls{F}/m_{\mu}})
+C_{P}^{2} (x-{m\ls{F}/m_\mu}) \big \}
\over m_{\mu}^{2} x^{2} +(\mhpm^{2}-m_{\mu}^{2})x+
m\ls{F}^{2}(1-x)}\,.$$
Our calcuation agrees with that of Leveille [6].  (In the results
of ref.~[6], one must set the charged Higgs boson charge equal to $-1$
in units of $e$, since the calculations presented there are
for $g-2$ of the $\mu^-$.)  In the two-Higgs doublet model (see
\S 4.1), $C_S$ and $C_P$ depend on the choice of Higgs-fermion
couplings.  The Model-II couplings [see eq.~(4.22) and the discussion
following] yield $C_S^2=C_P^2=G_F m_\mu^2\tan^2\beta/\sqrt{2}$.
Inserting these
couplings into the equation above
(and noting that $F$ is the neutrino), we obtain in the limit of
$\mhpm\gg m_\mu$
$$
\left.\Delta a_{\mu} \right|_{\rm fig.~3.7}
={-G_{F} m_{\mu}^{4} \tan^2\beta\over 24 \sqrt{2}\,\pi^{2}\mhpm^2}\,.
\eqno{(3.9)}$$
This result should replace eq.~(3.9) at the bottom of p.~96.
Note that the above
result disagrees with the results of Grifols and Pascual
(ref.~[8] of Chapter 3) and Donoghue and Li (ref.~[20] of Chapter 4)
both in magnitude (by a factor of 2) and in sign.  We have carefully
checked that the sign of $\Delta a_\mu$ in eq.~(3.9)
is indeed opposite to that of
the standard QED one-loop correction.  In addition,
Grifols and Pascual [8]
denote $\kappa\equiv\tan\beta$, but incorrectly state that
$\kappa\leq 1$.  Thus, the statement below eq.~(3.9) on p.~96 should
be eliminated.
\point
In view of the previous erratum, eq.~(3.11) on p.~98
is not correct.  It should read:
$$
{\Delta a_{\mu} |\ls{\rm fig.~3.8} \over
\Delta a_{\mu} |\ls{\rm fig.~3.7}}\simeq {77\over\tan^2\beta}\,.
\eqno{(3.11)}
$$
In obtaining this result, we have also made use of the most recent
Particle Data Book values for the fundamental constants.  Therefore,
the two-loop charged Higgs contribution is {\it larger} than the
the corresponding one-loop contribution unless $\tan\beta\gsim 9$.
\point
In chapter 3, there is a typographical error in some of the
headlines appearing on odd-numbered pages.  Headlines on
pp.~109, 111, 113, 115, and 129 should read ``Very Light Higgs
Bosons (mass $\lsim 5$~GeV)''.  Headlines on pp.~137 and  139 should
read ``Light Higgs Bosons ($5\lsim\mhsm\lsim 85$~GeV)''.
\point
On p.~155 in the line right above eq.~(3.94), the text should read:
``Before applying any cuts, reaction~(3.92) overtakes reaction~(3.93)
at roughly''.
\point
On p.~195, we state that
eq.~(4.8) is the most general two-Higgs doublet
scalar potential subject to a discrete symmetry $\phi_1\rta -\phi_1$
which is only softly violated by dimension-two terms.  This is not
strictly correct.  There is one additional term that can be
added:
$$\lambda_7\left[{\rm Re}~\phi_1^\dagger\phi_2-v_1 v_2\cos\xi\right]
\left[{\rm Im}~\phi_1^\dagger\phi_2-v_1 v_2\sin\xi\right]\,.$$
However, this term can be eliminated by redefining
the phases of the scalar fields.  To see this, note that if
$\lambda_7\neq 0$ then the coefficient multiplying
the term $(\phi_1^\dagger
\phi_2)^2$ in the scalar potential is complex, while if $\lambda_7=0$
then the corresponding coefficient is real.  Subsequent results
presented in \S 4.1 are not affected by this choice.  Moreover,
in the minimal supersymmetric model, $\lambda_7=0$ (at tree-level).
On the other
hand, in CP-violating two-Higgs-doublet models, it is important to
keep $\lambda_7\neq 0$ if one wishes to retain the overall freedom
to redefine the Higgs field phases.
\point
On p.~209, in the text five lines above eq.~(4.46), replace the
reference to eq.~(2.29) with eq.~(2.30).
\point
On p.~264, the reference to Witten in ref.~106
is incorrect.  It should read:
E. Witten, {\it Nucl. Phys.} {\bf B188} (1981) 513.
\point
On p.~264, there are some errors in ref.~107.  First, A. Kakuto's
name was mistakenly omitted.  Second, a reference to one paper
was accidently left out.  The corrected reference should read:
K. Inoue, A. Kakuto, A. Komatsu and S. Takeshita, {\it Prog. Theor.
Phys.} {\bf 67} (1982) 1889; {\bf 68} (1982)
927 [E: {\bf 70} (1983) 330]; {\bf 71} (1984) 413.
\point
Here is a clarification regarding the Feynman rules for cubic Higgs
vertices involving at least one Goldstone field, which appear
on pp.~373--375.  The rules for cubic and quartic scalar
interactions given in Appendix A are specific to the minimal
supersymmetric model (MSSM), since the MSSM imposes specific
constraints on the Higgs potential.  Nevertheless, it turns out
that the rules for cubic scalar interactions involving at least one
Goldstone boson can be written in a completely model independent
way.  Three examples of these model-independent rules (which
apply to the most general CP-invariant two-Higgs-doublet model) were
given in Fig.~A.17 on p.~375.  In fact, we can obtain similar
results for the other six non-zero three-Higgs vertices given
on pp.~373--374.  For completeness, the
Feynman rules in the general CP-invariant two-Higgs-doublet model
for all non-zero three-Higgs vertices involving
at least one Goldstone field are listed below:
$$\eqalign{
g\ls{h^0G^0G^0} &=
   {-ig\over 2\mw} \mhl^2\sin(\beta-\alpha)\,,\crr
g\ls{H^0G^0G^0} &=
   {-ig\over 2\mw} \mhh^2\cos(\beta-\alpha)\,,\crr
g\ls{h^0G^+G^-} &=g\ls{h^0G^0G^0}\,,\crr
g\ls{H^0G^+G^-} &=g\ls{H^0G^0G^0}\,,\crr
g\ls{h^0A^0G^0} &={-ig\over 2\mw}(\mhl^2-\mha^2)\cos(\beta-\alpha)\,,\crr
g\ls{H^0A^0G^0} &={ig\over 2\mw}(\mhh^2-\mha^2)\sin(\beta-\alpha)\,,\crr
g\ls{h^0H^\pm G^\mp} &=
{ig\over 2\mw}(\mhpm^2-\mhl^2)\cos(\beta-\alpha)\,,\crr
g\ls{H^0H^\pm G^\mp} &=
{-ig\over 2\mw}(\mhpm^2-\mhh^2)\sin(\beta-\alpha)\,,\crr
g\ls{A^0H^\pm G^\mp} &=
{\pm g\over 2\mw}(\mhpm^2-\mha^2)\,.\cr}$$
In the rule for the $\ha H^\pm G^\mp$ vertex, the sign corresponds to
$H^\pm$ entering the vertex and $G^\pm$ leaving the vertex.
[If CP were not conserved, these rules would still apply, except
that $h^0$, $H^0$ and $A^0$ would no longer be mass eigenstates.]
One can easily check that if tree-level MSSM relations are imposed
on the Higgs masses, and angles $\alpha$ and $\beta$ [see
eqs.~(A.9)--(A.12) on p.~355],
one recovers the MSSM Feynman rules listed on pp.~373--374 in
Figs.~A.15 and A.16.
\point
On p.~375, in the caption to Fig.~A.17, on the third line, replace
the reference to eq.~(A.10) with eq.~(A.9).
\point
On pp.~404--405, in Fig.~A.35(c) and (d),
the Feynman rules for the $H^0h^0\widetilde q_{kL}
\widetilde q_{kL}$ and $H^0h^0 \widetilde q_{kR}
\widetilde q_{kR}$ vertices are incorrect.  The correct Feynman
rules are:
$$\eqalign{\qquad H^0h^0\widetilde q_{kL}\widetilde q_{kL}
\qquad\qquad&{ig^2\sin2\alpha\over 2}\left[{T_{3k}-e_k\sin^2\theta_W
\over\cos^2\theta_W}-{m_q^2\over2m_W^2}D_k\right]\,,\crr
H^0h^0 \widetilde q_{kR}\widetilde q_{kR}
\qquad\qquad&{ig^2\sin2\alpha\over 2}\left[e_k\tan^2\theta_W
-{m_q^2\over2m_W^2}D_k\right]\,.\cr}$$
\point
On p.~407, in Fig.~A.36(f),
the Feynman rule for the $G^\pm H^\mp\widetilde q_{kR}
\widetilde q_{kR}$ vertex is incorrect.
The corrected rule is:
$$G^\pm H^\mp\widetilde q_{kR}\widetilde q_{kR}
\qquad\qquad{ig^2\over 2}\,\sin2\beta\,e_k\tan^2\theta_W
-{ig^2m_q^2\over2m_W^2}H_k\,.$$
\point
On p.~417, in eq.~(C.1), replace $\Gamma(\hsm\rta\gamma\gamma)$
with $\Gamma(h\rta\gamma\gamma)$, in order to be consistent
with the notational conventions of Table 1.1 on p.~4.
\point
On p.~418, in the second line below eq.~(C.3), it should read:
``$F_{1/2}^{A^0}\rta -2$''.
\point
On p.~421, in the index listing for ``Decays of $A^0$\dots\ to
$Zh$'', replace 413 with 412.
\point
On p.~422, in the index listing for ``Decays of neutral Higgs
bosons \dots\ to $W^+W^-$; $ZZ$'', replace 413 with 412.
\bye